\documentclass{emulateapj}
\usepackage{natbib,graphicx}

\def \aj {AJ}
\def \mnras {MNRAS}

\def \apj {ApJ}
\def \apjs {ApJS}
\def \apjl {ApJL}
\def \aap {A\&A}
\def \nat {Nature}
\def \araa {ARAA}

\newcommand{\kms} {$\mathrm{km\;s^{-1}}$\ }
\newcommand{\msun} {M$_{\odot}$}

\newcommand{\rsun} {R$_{\odot}$}

\def\lesssim{\mathrel{\hbox{\rlap{\hbox{\lower4pt\hbox{$\sim$}}}\hbox{$<$}}}}
\def\gtrsim{\mathrel{\hbox{\rlap{\hbox{\lower4pt\hbox{$\sim$}}}\hbox{$>$}}}}

\shorttitle{SN 2009kf}
\shortauthors{Botticella et al.}

\begin{document}


\title{SN 2009kf : a UV bright type IIP supernova discovered with
  Pan-STARRS 1 and GALEX}


\author{M. T. Botticella\altaffilmark{1}, 
              C. Trundle\altaffilmark{1}, 
              A. Pastorello \altaffilmark{1},
              S. Rodney\altaffilmark{2},
              A. Rest\altaffilmark{3},
              S. Gezari\altaffilmark{2},
              S. J. Smartt\altaffilmark{1},
              G. Narayan\altaffilmark{3},
               M. E. Huber\altaffilmark{2},
               J. L. Tonry\altaffilmark{5},              
                D.Young\altaffilmark{1}, 
                K. Smith\altaffilmark{1}, 
                F. Bresolin\altaffilmark{5},
                S. Valenti\altaffilmark{1},                 
                 R. Kotak\altaffilmark{1}, 
                S. Mattila\altaffilmark{6}, 
                E. Kankare\altaffilmark{7,6},  
                 W. M. Wood-Vasey\altaffilmark{8},  
                 A. Riess\altaffilmark{9},              
                  J. D. Neill\altaffilmark{10}, 
                  K. Forster\altaffilmark{10}, 
                  D. C. Martin\altaffilmark{10},
                   C.  W. Stubbs\altaffilmark{3}, 
                   W. S. Burgett\altaffilmark{5}, 
                  K. C. Chambers\altaffilmark{5}, 
                  T. Dombeck\altaffilmark{5}, 
                  H. Flewelling\altaffilmark{5},
                  T. Grav\altaffilmark{2}, 
                  J. N. Heasley\altaffilmark{5}, 
                  K.  W. Hodapp\altaffilmark{5}, 
                  N. Kaiser\altaffilmark{5}, 
                  R. Kudritzki\altaffilmark{5}, 
                  G. Luppino\altaffilmark{5},  
                   R. H. Lupton\altaffilmark{11}, 
                   E. A. Magnier\altaffilmark{5}, 
                   D. G. Monet \altaffilmark{12}, 
                   J. S. Morgan\altaffilmark{5}, 
                   P.  M. Onaka\altaffilmark{5}, 
                   P. A. Price\altaffilmark{5}, 
                   P. H. Rhoads\altaffilmark{5}, 
                   W. A. Siegmund\altaffilmark{5},
                   W. E. Sweeney\altaffilmark{5},
                   R. J. Wainscoat\altaffilmark{5}, 
                   C. Waters\altaffilmark{5}, 
                   M. F. Waterson\altaffilmark{5}, 
                   C.  G. Wynn-Williams\altaffilmark{5}}
\altaffiltext{1}{Astrophysics Research Centre, School of Maths and Physics, Queen's University, BT7 1NN, Belfast, UK}
\altaffiltext{2}{Department of Physics and Astronomy, Johns Hopkins University, 3400 N. Charles St., Baltimore, MD, 21218, USA}
\altaffiltext{3}{Department of Physics, Harvard University, Cambridge,  MA 02138, USA}
\altaffiltext{4}{Harvard-Smithsonian Center for Astrophysics, 60 Garden Street, Cambridge, MA 02138, USA}
\altaffiltext{5}{Institute for Astronomy, University of Hawaii at Manoa, Honolulu, HI 96822, USA}
\altaffiltext{6}{Tuorla Observatory, Department of Physics and Astronomy, University of Turku, Piikki\"o, FI 21500, Finland}
\altaffiltext{7}{Nordic Optical Telescope, Apartado 474, E-38700 Santa Cruz de La Palma, Spain}
\altaffiltext{8}{Department of Physics and Astronomy, University of Pittsburgh, Pittsburgh, PA 15260, USA}
\altaffiltext{9}{Space Telescope Science Institute, 3700 San Martin Drive, Baltimore, MD 21218-2463, USA}
\altaffiltext{10}{California Institute of Technology, 1200 East California Blvd., Pasadena, CA  91125, USA}
\altaffiltext{11}{Department of Astrophysical Sciences, Princeton University, Princeton, NJ 08544, USA}
\altaffiltext{12}{US Naval Observatory, Flagstaff Station,  Flagstaff,  AZ 86001, USA}

\begin{abstract}

We present  photometric and spectroscopic observations of a luminous type IIP Supernova (SN) 2009kf discovered 
by the Pan-STARRS 1 (PS1) survey and detected also by \textsl{GALEX}.   The SN shows a plateau in its optical 
and bolometric light curves, lasting approximately 70 days in the rest frame, with absolute magnitude of M$_V = -18.4$\, mag. 
The P-Cygni profiles of 
hydrogen indicate  expansion velocities  of 9000\,\kms  at  61 days after discovery which is extremely high for a type IIP SN. 
SN~2009kf is also remarkably bright in the near-ultraviolet (NUV) and shows a slow evolution 10--20 days after optical discovery.
The NUV and optical  luminosity at these epochs can be 
modelled with a black-body with a hot effective temperature  ($T\sim
16,000$\,K) and a large radius ($R\sim 1\times 10^{15}$\,cm). 
The bright  bolometric and NUV luminosity, the lightcurve peak and plateau duration, the high velocities and temperatures suggest that 2009kf is a type IIP SN  powered by
a larger than normal explosion energy.   
Recently discovered high-z SNe ($0.7 < z < 2.3$)
have been assumed to be IIn SNe, with the bright UV luminosities due to the interaction of SN ejecta with a 
dense circumstellar medium (CSM).  UV bright SNe similar to SN 2009kf could also account for these high-z events, and its absolute 
magnitude $M_{\rm NUV} = -21.5 \pm0.5$\,mag suggests such SNe could be discovered out to $z\sim2.5$ in the 
PS1 survey. 

\end{abstract}



\section{Introduction} 
\label{sec:intro}
Type II SNe are hydrogen rich explosions and fall into three main
sub-classes.  Type IIP  events have plateaus in their optical and
near-infrared (NIR) lightcurves,  type  IIL  events  show a linear
decay after peak,  type IIn events  present strong  
signatures of the presence of dense CSM and are characterized by 
narrow hyrdogen emission lines superimposed on broad wings. 
The relative fractions of these SNe are now well
measured in the nearby Universe
\cite[see][for a review of relative rates]{Smartt2009A}.  
The majority of these (around 60\%) are IIP with typical mid-plateau magnitudes
of $M_{V}\sim-17$\,mag 
\citep{Richardson2002}. 
However,  they are heterogeneous and span 
a factor of 100  both in luminosity and  in 
mass of $^{56}$Ni created explosively \citep{Hamuy2002,Pastorello2003}. 
The progenitor stars of several of the nearest IIP SNe have been
discovered  \citep{Li2006,Smartt2009} and are the red supergiants (RSG) that both stellar evolutionary theory and lightcurve
modelling have predicted.

Type IIL  and type IIn  SNe are significantly less frequent by volume
making up about 3\% and 4\% respectively of the total core-collapse  (CC) SNe
\citep{Smartt2009}.   
The lack of an extended plateau suggests that type IIL SNe have more massive progenitor stars  that shed a considerable amount of  their hydrogen envelope before explosion but there are not progenitor detections that confirm this scenario. The progenitors of type IIn SNe have likely undergone large mass ejection just before their explosion \citep{GalYam2007,GalYam2009}.

Type II  SNe are now being searched for in 
medium and high redshift surveys for a variety of reasons. 
The IIP SNe appear to be reasonably standard candles  and
have produced precise distance estimates with different methods \citep{2004ApJ...616L..91B,Dessart2008}, but  the empirical correlation of 
plateau-luminosity and expansion velocity, the standardised candle method (SCM),  seems a promising
method of measuring the distances to large numbers of  type IIP SNe. 
The dispersion of order 0.2--0.3\,mag \citep{Hamuy2003,Nugent2006,Poznanski2009},  0.1--0.15\,mag in NIR range\citep{Maguire2009b},
is potentially similar to type Ia SNe but they 
are significantly fainter,  restricting their use with current
surveys to $z<0.3$. 

Although this is
CC SNe of all varieties have been used to estimate the
star formation rate (SFR) out to $z\sim0.2$
\citep{Botticella2008,Bazin2009} and to $z\sim0.7$  \citep{Dahlen2004}. 
Recently the highest redshift  SNe ($0.7 < z < 2.3$) have been found, 
and proposed to be UV bright IIn SNe \citep{Cooke2009}. \citet{Cooke2008}
suggested that this type of SNe
could be used to probe the SFR of the  Universe out to $z\sim2$ in upcoming surveys. 

The PS1 survey has the potential to discover 
thousands of SNe between  $0 < z\le 1$ \citep{Young2008}.
The seven square degree camera and 1.8m 
aperture could allow IIP SNe to be used as cosmological probes at 
$z\sim0.2$ and the brightest  events to be found out to $z\sim2$.
One of the first discoveries of PS1, SN~2009kf \citep{2009CBET.1988....1Y}, is a very bright
SN that  shares some characteristics with IIP SNe with its 
luminous plateau  and  broad P-Cygni features. Simultaneous 
\textsl{GALEX} images show that it is also remarkably bright in the NUV. 
We discuss the implication 
of this rare SN for understanding the explosions and the use of 
type IIP events for probing cosmology and SFR at high redshifts. 
We adopt the cosmological parameters 
$H_0=70\,{\rm km}\,{\rm s}^{-1} {\rm Mpc}^{-1}$, $\Omega_M=0.3$, $\Omega_\Lambda=0.7$.  

\section{Discovery and observational data}\label{data}
\begin{figure}
\epsscale{1}
\plottwo{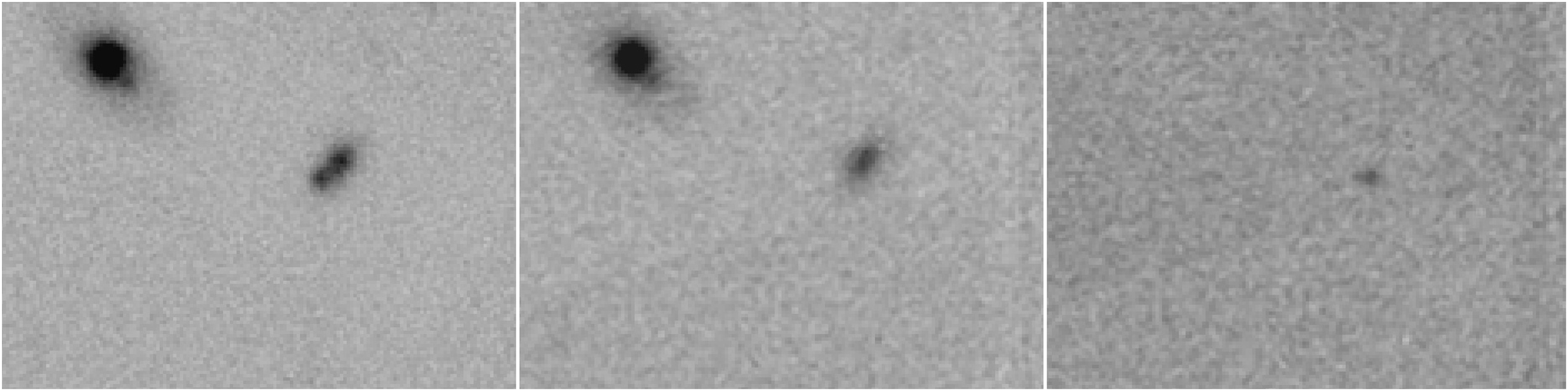}{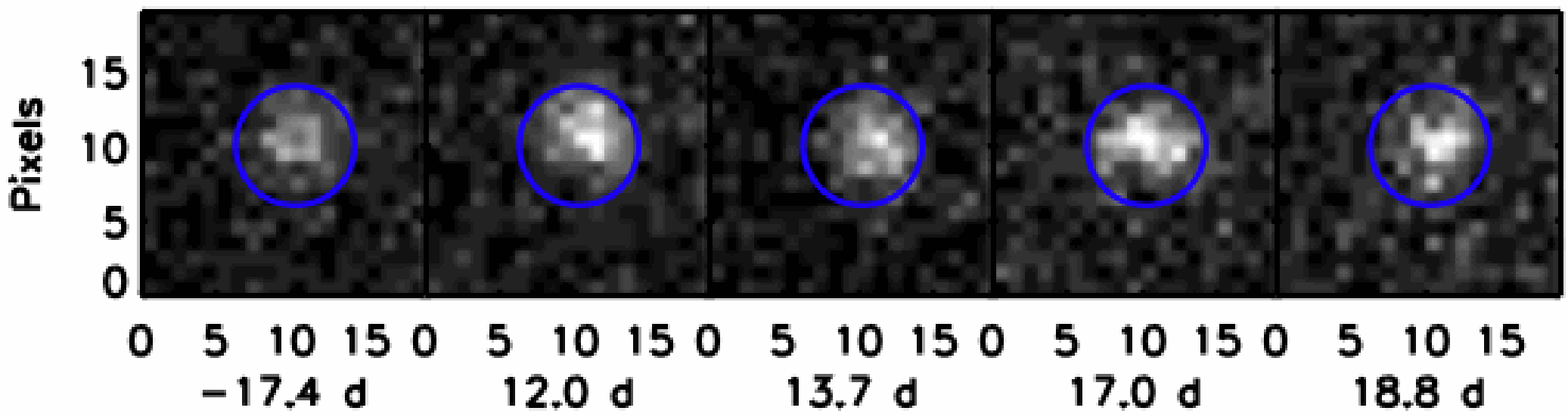}
\caption{Left panel:  Discovery image of SN 2009kf ($r$ band) in MD08 sky field, template  and difference images. 
Right Panel: \textsl{GALEX} NUV images of the host galaxy  
pre-SN and with SN 2009kf in rest-frame. The phase is with respect to the explosion date (JD$=2,454,989.5\pm 3$).  Blue circle shows the 6 arcsec (4 pixel) radius  
aperture used to measure the fluxes.
\label{figdisc}}
\end{figure}

SN~2009kf     ($\rm{\alpha_{J2000}=16^{h}12^{m}54^{s}.05,  \delta_{J2000}=+55\degr38\arcmin13\arcsec.7}$) was discovered  on  2009 Jun 10.9 UT by PS1  during the 
course of the  Medium Deep Survey (10 extragalactic
fields observed nightly using 25\% of the telescope time) in the sky-field MD08 (Fig.~\ref{figdisc}). 
The last  non detection was  on  2009 Jun~04.9  UT so we adopt  an explosion 
date ${\rm JD}=2,454,989.5\pm 3$.
The  reported phases are with respect to the explosion date and in the SN rest frame.
The PS1 MD08 coverage provided $g,r,i,z$ photometry, and we supplemented this with images from the 
Liverpool Telescope (LT), William Herschel Telescope (WHT) and Gemini-North Telescope (GN). 
The PS1 images were reduced with the custom built Image Processing Pipeline 
while LT, WHT and GN images were reduced with IRAF\footnote{IRAF is distributed by the National Optical Astronomy Observatories, which are operated by the Association of Universities for Research in Astronomy, Inc., under cooperative agreement with the
National Science Foundation.} tasks.
The instrumental magnitudes  were derived from host galaxy template subtracted images using PSF fitting techniques \citep[as in][]{Botticella2009}.
A local sequence of SDSS stars was used to measure the relative magnitude for each observation.
The uncertainties in the measurements  are estimated by combining in quadrature the error
of the photometric calibration and the error in the PSF fitting. 
Apparent and absolute magnitudes reported in this letter are in the AB and Vega system, respectively.

The \textsl{GALEX} Time Domain Survey (TDS) detected SN~2009kf  in
monitoring observations of the MD08 field.  The host galaxy was
detected in the pre-explosion TDS observations between ${\rm JD}=2,454,960.7$ to 2,454,968.9
with ${\rm NUV }= 21.58 \pm 0.15$\,mag. This 
is corrected for the flux enclosed in a 6\,\arcsec radius aperture
\citep{Morrissey2007} and the error is measured from the dispersion of measurements between the observations. 
The SN was detected in 5 observations with exposure times  of  $1.0-1.5$\,ks
 from ${\rm JD}=2,455,003.7$ to 2,455,013.8, with a peak magnitude of
${\rm NUV }=22.34 \pm 0.38$\,mag. The SN magnitude was measured by subtracting the
flux of the host galaxy from the total flux enclosed in a 6\,\arcsec radius aperture centered on the 
host galaxy NUV centroid as determined in the pre-discovery images. 
The errors in the SN
magnitude include both the error in the host galaxy magnitude as well
as the error in the observed magnitude, which is measured empirically as a
function of magnitude from the dispersion of 4275 matched sources
 between observations in bins of 0.5\,mag.

Spectroscopic follow-up was obtained with Gemini Multi-Object Spectrographs (GMOS) at GN  on ${\rm JD}=2,455,061.7$ and 2,455,094.7  and with Andalucia Faint Object Spectrograph and Camera (ALFOSC) at the Nordic Optical Telescope (NOT) on ${\rm JD}=2,455,117.6$. 
Spectra were reduced  using the  Gemini pipeline and standard routines within IRAF for the NOT spectrum.  
The spectra of the host galaxy SDSS $J161254.19+553814.4$ provided a
redshift measurement of $z=0.182 \pm 0.002$  from the nebular emission lines, 
in agreement with the SDSS photometric redshift of $0.185 \pm 0.065$\footnote{Information on the host galaxy is available at http://cas.sdss.org/astrodr7}.   
A reddening coefficient of C(${\rm H}\beta)=1.3\pm 0.25$ 
was  derived from  the ${\rm H}\alpha/{\rm H}\beta$ emission line ratio,  
which corresponds to an host galaxy extinction of $E(B-V)=0.9$\,mag,  while the Galactic extinction is negligible ($E(B-V)=0.009$\,mag \citet{Schlegel1998}).  The N\,{\sc ii} and O\,{\sc iii}N\,{\sc ii} line ratio methods of 
estimating  metallicity were employed  \citep{Pettini2004}, giving $12+\log ({\rm O/H})$ abundance of  $8.50 \pm 0.1$ dex.

\section{Photometric  and spectroscopic evolution}\label{analysis}
\begin{figure}
\epsscale{1}
\plotone{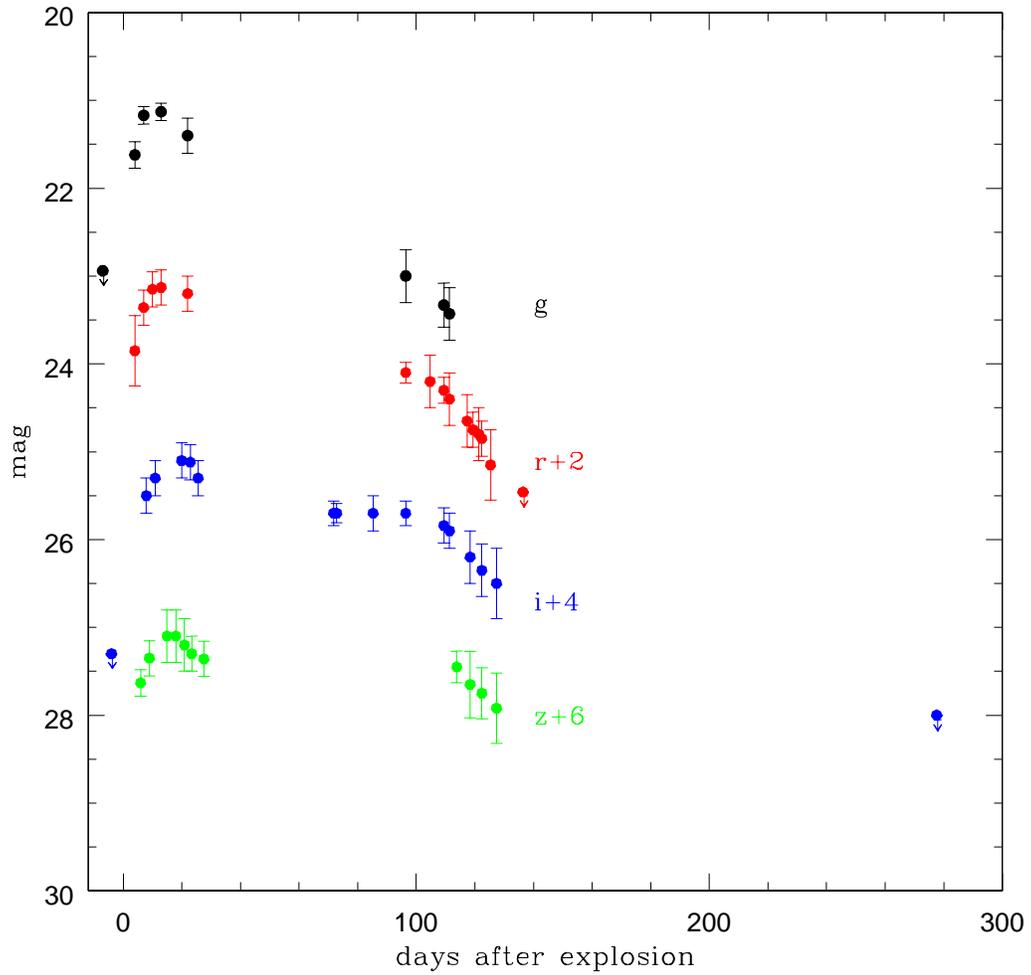}
\caption{The $g,r,i,z$ light curves of SN 2009kf in the observer frame. The magnitudes are not corrected for extinction. The phase is with respect to the explosion date. \label{fig1}}
\end{figure}

The $r,i,z$ lightcurves of SN~2009kf (Fig.~\ref{fig1}) display a
plateau  which is similar to that observed in  type IIP SNe \citep{Hamuy2003}. 
However there are differences in that SN~2009kf  shows a relatively slow rise to peak, 
and a clear maximum, which
occurs progressively earlier from the red to the blue bands.   In the last observation at about 280 days after explosion we estimated an upper limit of $i > 24$\,mag.

To meaningfully compare the  lightcurves  of SN~2009kf with those of  nearby type IIP SNe we estimated extinction and redshift corrections,  the latter requiring time dilation correction and K-correction.
We adopted a redshift of $0.182\pm0.002$ ($\mu=39.76$\,mag) and determined the K-correction from observed $riz$ AB magnitudes to rest frame $VRI$  Vega
magnitudes using the spectra of SN~2009kf, a sample of  spectra 
of different type II SNe and employing the {\em IRAF} package {\em synphot}.

\begin{figure}
\epsscale{1}
\plotone{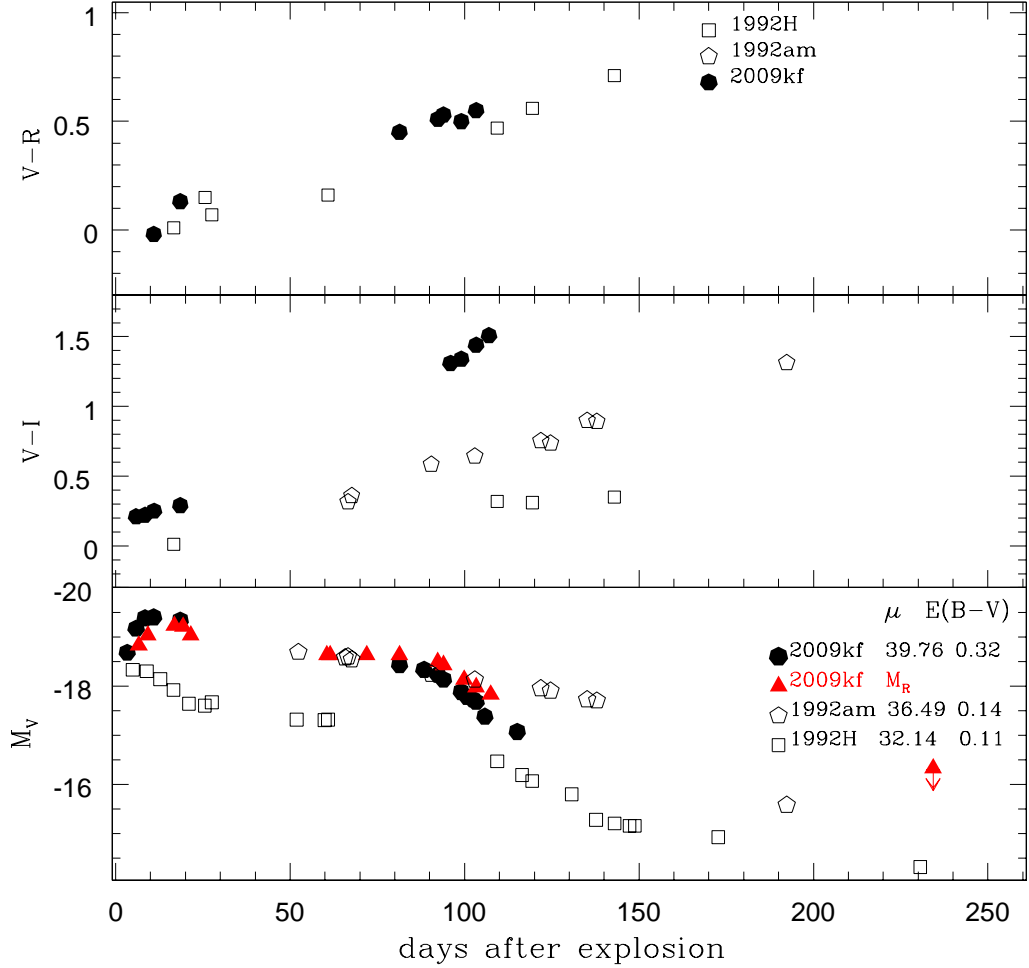}
\caption{Top and middle panel:  temporal evolution of rest frame $V-R$ and $V-I$ colours of SNe 2009kf, 1992H  
\citep{Clocchiatti1996} and 1992am \citep{Schmidt1994,Hamuy2003}. The colours of SN 2009kf are not corrected for host galaxy extinction while the colours of SNe 1992H and 1992am have been corrected.  The phases in the rest frame are relative to explosion epoch.
In the bottom panel the $V$ band absolute lightcurves of SNe 2009kf , 1992H and 1992am have been compared.  The absolute magnitudes have been corrected by internal and Galactic extinction. The phases in the rest frame are relative to explosion epoch. The $R$ band  absolute lightcurve of SN 2009kf is also illustrated.\label{abscomp}}
\end{figure}

The comparison of  colour evolution with that  of similar  SNe for which the colour excess has been 
previously determined can be used to estimate the extinction correction.  Fig.~\ref{abscomp} shows  the evolution of the $V-R$ and  $V-I$ colours not corrected for internal extinction  for SN~2009kf and of  the intrinsic colours  for  SNe 1992H and  1992am  \citep{Clocchiatti1996,Schmidt1994,Hamuy2003}.   
The intrinsic colour $V-I$ of type IIP SNe in the plateau phase appears fairly homogeneous,  
due to the photospheric temperature being close to that of hydrogen recombination. 
\citet{Olivares2010} suggest that this temperature leads to an intrinsic $(V-I)_0  = 0.66\pm0.05$\,mag at  the end of the plateu phase  and that $E(V-I)$ can be used to determine the extinction correction. 
SN~2009kf  has  $V-I= 1.3\pm0.4$\,mag at this epoch which implies a  value of $A_V = 1.6\pm1$\,mag.
We also compared the $V-R$ colour of SN~2009kf with that of SN~1992H  \citep{Clocchiatti1996}  since these SNe  are
similar both in the photometric and spectroscopic evolution.
We measured, at the same epoch, $V-R=0.55\pm 0.4$\,mag for SN~2009kf and  $V-R~0.25$\,mag for SN~1992H, 
which implies a value of $A_V=1\pm 1.4$\,mag.
The absorption component of the Na\,{\sc i}  doublet ($\lambda\lambda
5890,5896$) from the host galaxy in the spectra of SN~2009kf is not
detected, and we set a upper limit of EW(Na\,{\sc i} D) $< 1.4$\,\AA. 
The relation by \cite{Munari1997} then gives an upper limit on extinction of 
$E(B-V) \lesssim 0.3$\,mag. 
The integrated host galaxy extinction, as measured from the
$c({\rm H}\beta$) index is  $E(B-V)=0.9$\,mag. While this is not directly applicable to the 
the line of sight of SN~2009kf, it does suggest areas of high extinction are plausible. 
While the uncertainties prevent a definitive and consistent determination of reddening, 
the $VRI$ colors of 2009kf suggest a reddening  of $A_V \simeq 1$\, mag ($E(B-V)=0.32\pm 0.5$\,mag assuming $R_V=3.1$) is applicable. 

\begin{figure}
\epsscale{1}
\plotone{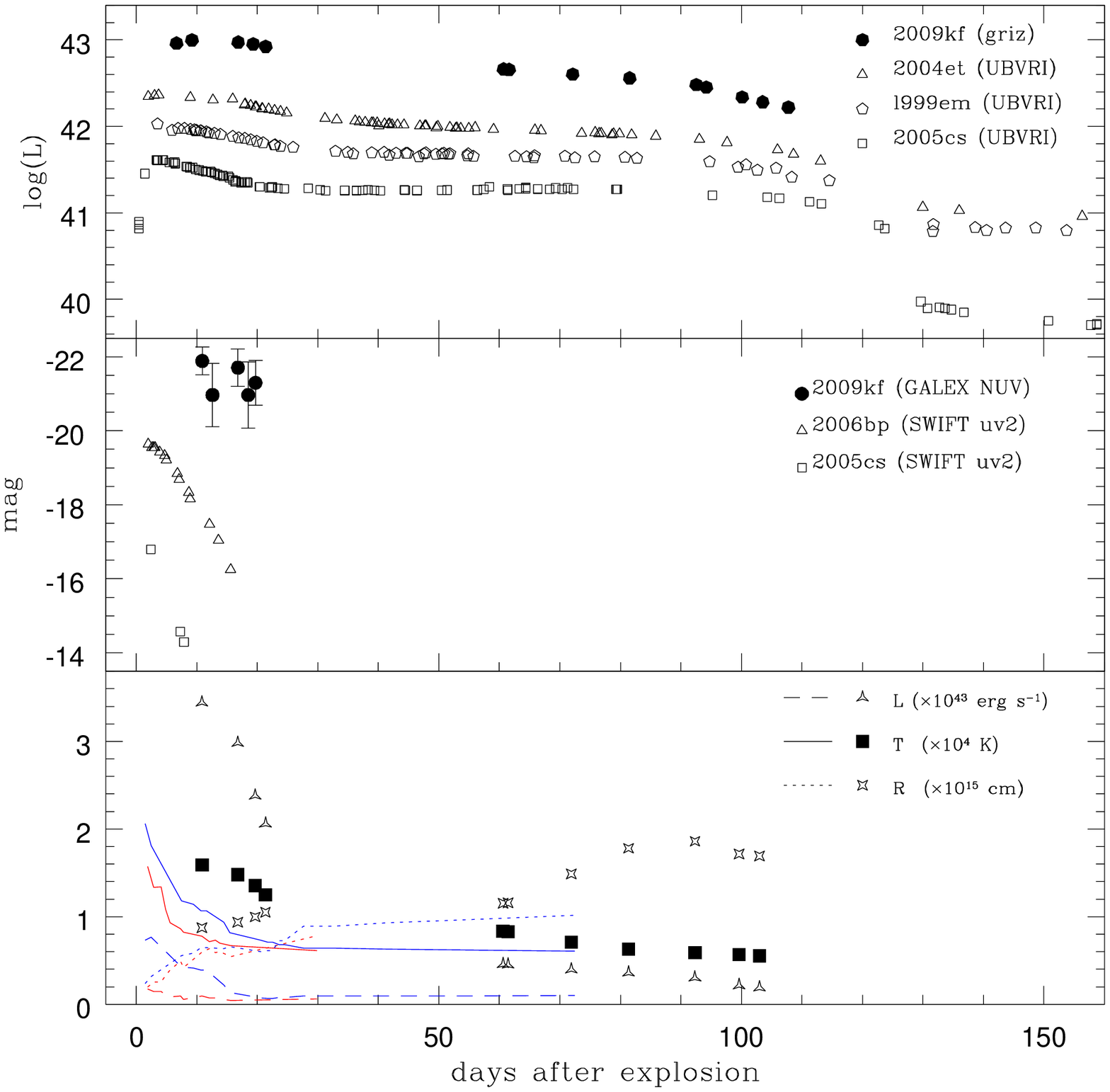}
\caption{Top panel: Psuedo bolometric light curves of SNe 2009kf,  2004et \citep{Maguire2009a},  1999em\citep{Hamuy2001}, 2005cs \citep{Pastorello2009}. The light curves are in the SN rest frames and the phases  are relative to explosion epoch.
Bottom Panel:  Absolute UV light curves of SNe 2009kf, 2006bp and
2005cs \citep{Dessart2008}. The light curves are in the SN rest frames
and the phases are relative to explosion epoch. The NUV magnitudes of
SN 2009kf are in Vega mags and not K-corrected. The central wavelength of NUV filter in the SN rest frame is about 1960\,\AA\. Bottom panel: temporal evolution of the temperature, radius and luminosity of the black body fit for SNe 2009kf (black), 2006bp (blue) and 2005cs (red) \citep{Dessart2008}. \label{bolUV}}
\end{figure}

The de-reddened  $V$ and $R$ band absolute light curves  are illustrated in Fig.~\ref{abscomp}.
In Fig.~\ref{bolUV}  the  pseudo-bolometric  and absolute NUV lightcurves of SN~2009kf  are compared with those of other type IIP SNe. 
The pseudo-bolometric lightcurve was obtained by first converting $,g,r,i,z$ magnitudes
into monochromatic fluxes, then correcting these fluxes for the adopted extinction  according to the law from \citet{Cardelli1989}, and finally  integrating the resulting SED over  wavelength, assuming zero flux at the integration limits. The pseudo-bolometric lightcurve clearly displays a plateau, from day 20 to day 90, which is shorter than the 100--120 days typical of IIP SNe. 
However,  it is significantly more luminous than
all other IIP SNe for which a good estimate of bolometric luminosity is available. 
The upper limit luminosity estimated at about  day 280 suggests that the mass of radioactive $^{56}$Ni deposited in the ejecta  is $M{\rm (^{56}Ni)}< 0.4$ \msun.
The absolute NUV lightcurve is obtained correcting the \textsl{GALEX}  data for the redshift and extinction discussed above. The absolute magnitude is incredibly bright at 10--20 days after explosion
in comparison with other well studied IIP SNe \citep{Dessart2008}.
The NUV flux and optical emission 
can be reproduced satisfactorily with a blackbody fit with a temperature
dropping from  
$\sim$\,16,000\,K to $\sim$\,13,000\,K from 10 to 20 days (as illustrated in Fig\,\ref{bolUV}).
The radius of the blackbody fit  implies an expansion velocity of about 10000\,\kms, which is similar to 
the measured velocities from absorption lines. 
The evolution of the temperature, radius and  luminosity of the blackbody fit is also  different from normal type IIP SNe  that are characterized by  smaller values   and different declines. 
We note that it would also be possible to reproduce the SED with 
low extinction and a temperature of $T_{\rm eff} \sim 9600$\,K, but this would then require an expansion
velocity of about 12,000\kms maintained to 20\,days, unusually high for a normal IIP.

\begin{figure}
\epsscale{1}
\plotone{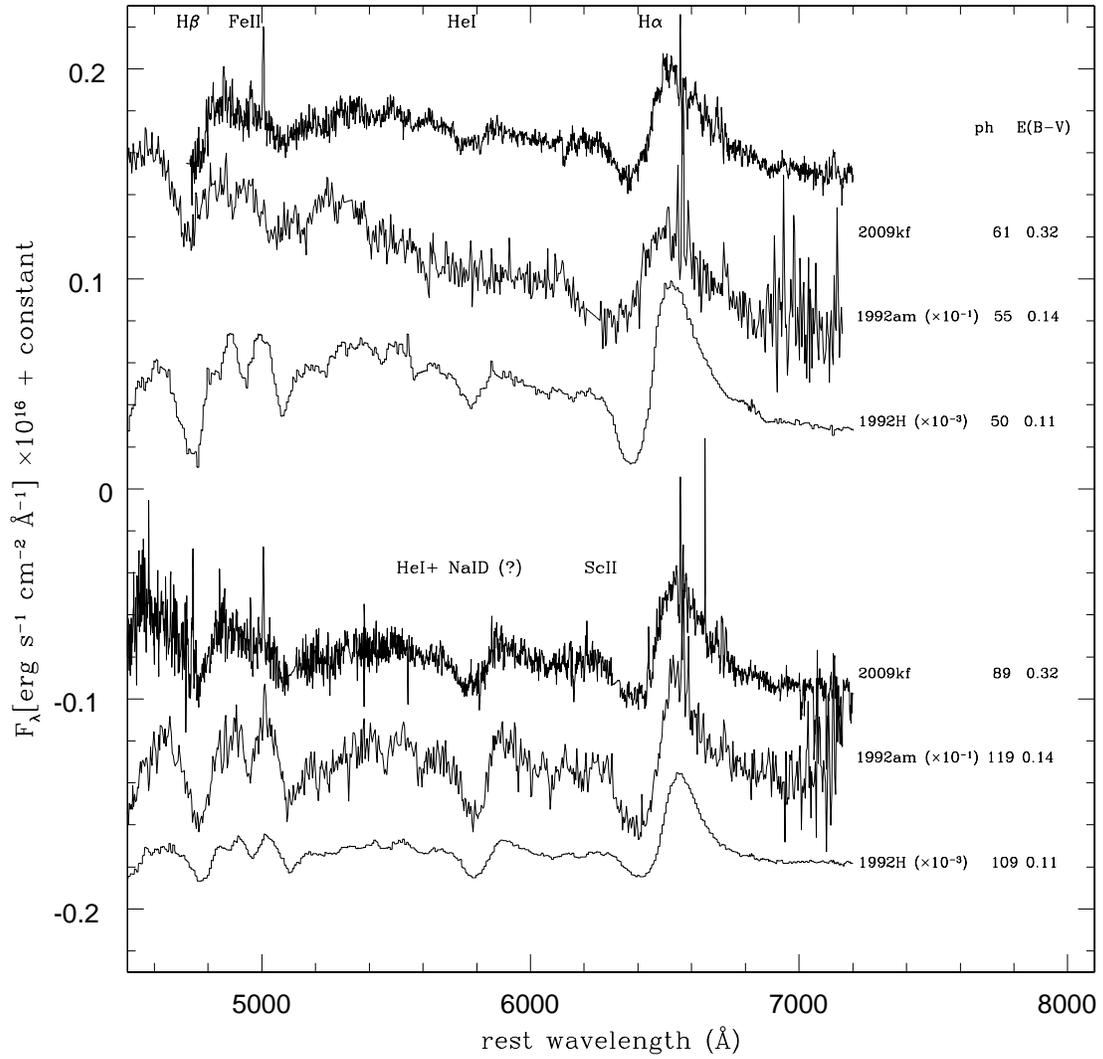}
\caption{Spectra of SNe 2009kf,  1992H \citep{Clocchiatti1996} and 1992am \citep{Schmidt1994,Hamuy2003} at two different epochs. The spectra  have been corrected for host galaxy recession velocities and for  reddening. The phases  in the rest frame are relative to explosion epochs.  \label{speccomp}}
\end{figure}

The spectra of SN~2009kf show strong ${\rm H}\alpha$ and  ${\rm H}\beta$ P-Cygni profiles, and a P-Cygni feature in the region around 5800\,\AA\ that may be attributed to He\,{\sc i}  ($\lambda$5876), Na\,{\sc i} ($\lambda$5889,5896) or their blend. However,  the peak of the emission is blue-shifted from the rest wavelength of He\,{\sc i}  by $100\pm 300$\,\kms and from that of Na\,{\sc i}  by $1000 \pm 300$\,\kms.  
We suggest that this is He\,{\sc i} since there are 
no prominent metal lines in the spectrum.  The presence of He\,{\sc i} is consistent with the high energy and  temperature inferred  by the NUV/optical lightcurves assuming an extinction of about $A_V=1$\,mag.
 Between 61 and 89 days after explosion, the continuum becomes redder  (Fig.~\ref{speccomp}) and there is little evolution in the Balmer features. Fitting Gaussian profiles to these features we determine expansion velocities of $9000 \pm 1000$\,\kms from ${\rm H}\alpha$ on day 61, and $7800  \pm 1000$\,\kms from ${\rm H}\alpha$ and ${\rm H}\beta$ on day 89 after discovery.  There is more evolution in He\,{\sc i}, which becomes stronger by day 89  due to an increasing contribution of  Na\,{\sc i}. The expansion velocity from this line decreases from $7200 \pm 1000$\,\kms on day 61 to $6500 \pm 1000$\,\kms on day 89. The spectra of SN~2009kf show similarities to SNe 1992H  and 1992am  \citep{Clocchiatti1996,Schmidt1994,Hamuy2003}(see Fig~\ref{speccomp}). 

\section{Discussion}\label{discussion}

The high luminosity, both in the optical and UV, short plateau duration and large expansion velocity of SN~2009kf have important implications for understanding the origins of type IIP SNe,  using them as cosmological distance indicators and detecting CC SNe at high redshift. 

The bolometric lightcurve of SN~2009kf  is similar to normal IIP SN but with three striking differences: it has a slow rising peak in the first 20 days, has a short plateau phase and is significantly more luminous. 
The NUV luminosity  is also remarkably higher and exhibits a slower evolution with respect to normal type IIP SNe \citep{2008ApJ...685L.117G}. 
The source of the observed luminosity  at  10 days  after the explosion can be modelled with a 
hot photosphere of $T_{eff}  \sim 16,000$\,K and a radius of around $1\times10^{15}$\,cm. 
The photospheric temperature remains high for an unusually long  period, as we see He\,{\sc i} ($\lambda$ 5876) up to about 89 days and the normal metal lines are weak. The expansion velocity as measured from the ${\rm H}\beta$ absorption is  similar to SN 1992am  and extreme for a IIP type event, 7800\,\kms at 89 days after discovery, compared to 4500\,\kms for a 
typical IIP.  
In conclusion 2009kf is both luminous and  extremely  blue hence  
extends the luminosity and energy range of IIP SNe.

This peculiarity raises the question of whether the standard model for type IIP SNe with a RSG progenitor of $8-20$\,\msun\,  and  an explosion energy of  $0.1-2 \times 10^{51}$\,ergs \citep{2009A&A...506..829U,Maguire2009a} is valid for this event.  
The luminosity, plateau duration and expansion velocity  mainly depend on the explosion 
energy, envelope mass and  radius of the progenitor star at the moment of the explosion and the characteristics 
 of SN~2009kf could be explained with a large explosion energy or very large progenitor radius, but a lower than usual hydrogen envelope mass. The analytical models of \citet{Kasen2009}  and numerical simulations of 
\citet{2004ApJ...617.1233Y} would require explosion energies in excess of   $10 \times 10^{51}$\,ergs, 
or progenitor radii of greater than 1000\,\rsun. 
Alternative explanations could be  a different mass distribution of H and He in the envelope, or possibly interaction of the ejecta with a surrounding shell. 
A detailed model of SN~2009kf  has the potential to provide insight  into the diversity of type IIP SNe, in particular as concern their use as standard candles.

The bright visual magnitude of SN~2009kf and its apparent plateau means that 
such events may be preferentially selected  in cosmology surveys. 
 However the distance modulus obtained  from the $I$-band relation by
 \citet{Nugent2006}  is $\mu=40.63 \pm 0.5$, compared to the distance
 modulus
of $\mu=39.76$ for a $\Lambda$CDM cosmology. Further work on these
high luminosity and fast expansion velocity events are required to
understand if this discrepancy for one event is statistical, or
systematic and if they can be reliably used for the SCM method applied to IIP SNe. 

The discovery of SN~2009kf  demonstrates the exciting potential of the  
\textsl{GALEX} TDS observing campaign which is coordinated with PS1 to  
both probe shock breakout \citep{2008ApJ...683L.131G,2008Sci...321..223S}
and  the nature of  UV-bright type II SNe.  During the first PS1 phases, three confirmed  SNe discoveries had
simultaneous \textsl{GALEX} imaging within $\pm$ 10 days of the PS1
discovery, but only SN 2009kf was detected. 
 The transients  discovered at high-z  by
\cite{Cooke2009} were interpreted as type IIn SNe, as up until now 
the only SNe which had such high UV luminosities were interacting events. 
SNe  2008es and 2009kf  are  the brightest SN in the UV known so  
far, ($M_{NUV}= -22.2$ and $ -21.5$\,mag respectively)  
but SN 2008es  \citep{Gezari2009, Miller2009} is a  type IIL SN and
did not have   obvious signs of CSM interaction. SN 2009kf is of
similar brightness to the inferred NUV fluxes of  \citet{Cooke2009} high-z 
SNe. 
The NUV lightcurves of all types of
CC SNe are not well quantified, hence it is quite possible that some fraction
of the  \citet{Cooke2009} high-z SNe are UV bright type II SNe similar to 2009kf. 
The seasonal stacked PS1 MDS images will allow 2009kf-like UV bright SNe to be detected beyond 
$z\sim2$ where the  NUV band would be 
redshifted to the $r$-band  and  use these 
events to probe the SF history of the high-z Universe. 
The rate of these SNe at low redshift, their progenitor scenarios, and 
their UV evolution are key areas we need to understand before we can confidently use
them to probe the SFR at  high redshift.


   \acknowledgments
The PS1 Surveys have been made possible through contributions of the  
Institute for Astronomy at the University of
HawaiÕi in Manoa, the Pan-STARRS Project Office, the Max-Planck  
Society and its participating institutes, the Max
Planck Institute for Astronomy, Heidelberg and the Max Planck  
Institute for Extraterrestrial Physics, Garching, The Johns
Hopkins University, the University of Durham, the University of  
Edinburgh, the QueenÕs University Belfast, the Harvard-
Smithsonian Center for Astrophysics, and the Los Cumbres Observatory  
Global Telescope Network, Incorporated.
This work is  also based on observations collected at LT, WHT and  NOT (La Palma) and Gemini (HawaiÔi).
This work, conducted as part of the award "Understanding the lives of
massive stars from birth to supernovae" (S.J. Smartt) made under the
European Heads of Research Councils and European Science Foundation
EURYI Awards scheme, see www.esf.org/euryi. M.T.B. would like to thank E. Cappellaro, L. Zampieri  and S. Benetti for helpful discussions.
SM and EK acknowledge support from the Academy of 
Finland (project:8120503).


\bibliographystyle{apj}

\clearpage

\end{document}